\documentstyle{article}

\newcounter{appendixc}
\renewcommand{\appendix}[1]
    {\addtocounter{appendixc}{1}
     \setcounter{equation}{0}
     \renewcommand{\theequation}{\Alph{appendixc}.\arabic{equation}}}

\begin{document}

\title{A Reconciliation of\\Collision Theory and
Transition State Theory}

\author{Yong-Gwan Yi}

\maketitle

\begin{abstract}
A statistical-mechanical treatment of collision leads to a formal
connection with transition-state theory, suggesting that
collision theory and transition-state theory might be joined
ultimately as a collision induced transition state theory.
\end{abstract}

\bigskip

Reaction rate theories are divided into two approaches $-$
collision theory and transition state theory.
Differences between the two theories and attempts to either
show their equivalence or to unite them into one have been
a subject of discussions in the literatures and textbooks [1].
In this paper, I should like to give a discussion of possible
joining of the two theories. Actually, whether implicit or explicit,
we have met their relationship in unimolecular reactions.
It is not my purpose to stress such a relationship,
but rather to present a general theory which unifies the two theories
and which may provide a framework of describing reaction rates.
The discussion goes back to the very beginnings of the reaction
rate theories.

Consider a collision process between two molecules of $A$
and $B$. We can discuss in the coordinate system of the
center of mass the collision that occurs between the two
molecules. All of the energy which goes into exciting the
activated complex must come from the energy of relative
motion of the reactants. Energy in the center of mass motion
cannot contribute. According to the kinetic theory of collision,
the rate constant has to be weighted by the Maxwell-Boltzmann
distribution function $f(u)$ of relative speed $u$, with
integration over speeds from zero to infinity, to give the
overall average rate constant:
\begin{equation}
k_C=\int_0^{\infty}\sigma u f(u) \, du,
\end{equation}
where $\sigma$ is the collision cross section. The rate constant
in this expression is given by M. Trautz in 1916 and by W. C. M.
Lewis in 1918.

It is instructive to evaluate the rate constant in terms of energy
states instead of direct integration. We now consider the basic
method of statistical mechanics of evaluating partition function [2].
Statistical mechanics states:

\begin{quotation}
\noindent The partition function is a sum over all states
$\Omega$, very many of which have the same energy.
One can perform the sum by first summing over all the
$\Omega(E)$ states in the energy range between $E$ and
$E+\delta E$, and then summing over all such possible energy
ranges. Thus
\begin{equation}
Q=\sum_n e^{-E_n/k_BT}=\sum_E \Omega(E)e^{-E/k_BT}.
\end{equation}
The summand here is just proportional to the probability that
the system has an energy between $E$ and $E+\delta E$.
Since $\Omega(E)$ increases very rapidly while
$\exp(-E/k_BT)$ decreases very rapidly with increasing $E$,
the summand $\Omega(E)\exp(-E/k_BT)$ exhibits a very
sharp maximum at some value $E^*$ of the energy. The mean
value of the energy must then be equal to $E^*$, and the
summand is only appreciable in some narrow range
$\Delta E^*$ surrounding $E^*$. The partition function must
be equal to the value $\Omega(E^*)\exp(-E^*/k_BT)$ of
the summand at its maximum multiplied by a number of the
order of $\Delta E^*/\delta E$, this being the number of
energy intervals $\delta E$ contained in the range $\Delta E^*$.
Thus
\begin{equation}
Q=\Omega(E^*)e^{-E^*/k_BT}
\biggl(\frac{\Delta E^*}{\delta E}\biggr),
\ \mbox{so}\ \ln Q=\ln\Omega(E^*)-
\frac{E^*}{k_BT}+\ln\biggl(\frac{\Delta E^*}{\delta E}\biggr).
\end{equation}
But, if the system has $f$ degrees of freedom, the last term
on the right is at most of the order of $\ln f$ and is thus utterly
negligible compared to the other terms which are of the order of
$f$. Hence, the result agrees with the general definition
$S=k_B\ln\Omega(E^*)$ for the entropy of a macroscopic
system of mean energy $E^*$.
\end{quotation}

We have seen the basic method of statistical mechanics of
evaluating the partition function. If we apply this to the
integration of Eq. (1), we would expect an expression for
the rate constant to be
\begin{equation}
k_C=\sigma u^*\biggl(\frac{\Delta E^*}{\delta E}\biggr)
\Omega(E^*) e^{-E^*/k_BT},
\end{equation}
where $u^*$ is a relative velocity for reaching the activated state.
This summation indicates that the integration over the translational
energy has a very sharp maximum at the activation energy $E^*$.
The width $\Delta E^*$ of the maximum, given by the square
root of the dispersion, is very small relative to $E^*$ for a
macroscopic system. The Maxwell-Boltzmann distribution function
we have used is the one normalized to unity on integration over
all states. For the results of more realistic calculation the
normalization should be expressed in its explicit form. The
following expression is then obtained:
\begin{equation}
k_C=\sigma u^*\biggl(\frac{\Delta E^*}{\delta E}\biggr)
\biggl(\frac{\Omega(E^*)}{Q_AQ_B}\biggr)e^{-E^*/k_BT}.
\end{equation}

This equation may equally be written in terms of an entropy change
in reaching the activated state. As $(\Delta E^*/\delta E)\Omega(E^*)$
represents a number of energy states in the activated state, the
expression for the rate constant may also be written in the form
\begin{equation}
k_C=\sigma u^* e^{\Delta S^*/k_B}e^{-E^*/k_BT}.
\end{equation}
The basic method of statistical mechanics shows how an entropy
term can be introduced in the kinetic theory expression. Here
$\Delta S^*$ represents the change in entropy due to the change
in energy in reaching the activated state. In a system of chemical
reaction the entropy of the system is a function of energy $E$,
volume $V$, and the number of molecules $N$: $S=S(E,V,N)$. Hence,
we can replace $\Delta S^*$ in Eq. (6), using the thermodynamic
relations, by its generalization
\begin{equation}
\Delta S^* \longrightarrow \Delta S^* - \frac{P\Delta V^*}{T}
+\frac{\mu\Delta N^*}{T},
\end{equation}
for a system of chemical reaction. The rate constant can then be
written in a general form
\begin{equation}
k_C=\sigma u^*e^{-\Delta G^*/k_BT},
\end{equation}
where $\Delta G^*$ is the Gibbs energy change in going
from the initial to the activated state.

We can now uncover
the collision theory expression for the rate constant of great
interest. It has been stated that collision theory of reaction
rates is not consistent with the fact that at equilibrium the
ratio of rates in the forward and reverse directions is
the equilibrium constant. However, it becomes evident that
the kinetic theory of collision does not lack the entropy term
that should appear in the expression for the equilibrium constant.
Collision theory of reaction rates is consistent with the fact.
Furthermore, it provides us with a kinetic theoretical derivation
of thermodynamic expression for the equilibrium constant, by setting
$\sigma u^*$ equal to that in the reverse direction according to
the principle of detailed balance. We were able to show
the thermodynamic equilibrium.

The transition-state theory was published almost simultaneously
by H. Eyring and by M. G. Evans and M. Polanyi in 1935. The rate
equation for a bimolecular reaction derived by this theory
of reaction is
\begin{equation}
k_{TS}=\frac{k_BT}{h}
\biggl(\frac{Q_{\ddagger}}{Q_AQ_B}\biggr)e^{-E^*/k_BT}
\end{equation}
from a partition function for the activated complex
$$
Q^{\ddagger}\approx \frac{k_BT}{h\nu}Q_{\ddagger}.
$$
The partition functions $Q_A$ and $Q_B$ relate to the two reactants,
and $Q_{\ddagger}$ is a special type of partition function for the
activated complex. It is just like a partition function for a normal
molecule, except that one of its vibrational degrees of freedom is
in the act of passing over to the translation along the reaction
coordinate.

Equation (5) is very suggestive in relating collision theory to
transition-state theory. The kinetic theory expression leads us
to an idea of connecting with transition-state theory formula.
By identifying $Q^{\ddagger}$ with $(\Delta E^*/\delta E)\Omega(E^*)$,
we can put both theories into some perspective. From their formal
connection the reaction can be viewed as a succession of two steps
$-$ collision and transition state. The overall rate is then given
by the sum of two average lifetimes:
rate $=(k_C^{-1}+k_{TS}^{-1})^{-1}$.
The rate reads explicitly
\begin{equation}
\mbox{rate}=\biggl[\biggl\{\sigma u^*
\biggl(\frac{Q^{\ddagger}}{Q_AQ_B}\biggr)\biggr\}^{-1}
+\quad\biggl\{\frac{k_BT}{h}\biggl(\frac{Q_{\ddagger}}
{Q_AQ_B}\biggr)\biggr\}^{-1}\biggr]^{-1}e^{-E^*/k_BT}.
\end{equation}

The transition-state theory is concerned with a motion in
the activated state and gives no explanation for reaching the
activated state. The transition-state theory itself cannot be
a complete treatment of reaction rates. A chemical reaction
starts with an external interaction for reaction to occur first.
It is very natural to consider molecular collisions as for
an external interaction. The simple kinetic theory counts
every sufficiently energetic collision as an effective one.
Equation (10) suggests correcting the collision frequency
by involving the translation along the reaction coordinate
also in the evaluation of the transition over translational
energy states. The essential feature of the argument is that
transition state is brought about by energetic collisions and
that the rate of a reaction is determined by the frequency of
these collisions and by the resulting translations along the
reaction coordinate.

Indeed, the present treatment of reaction rates reflects the most
important aspects of unimolecular reactions [3]. Equation (10) is
in exact agreement in form with the rate equation given by
Rice-Ramsperger-Kassel-Marcus (RRKM)  theory of unimolecular
reactions. The distribution function
that has been used in RRKM theory is equal in expression to that
given by the basic method of statistical mechanics of evaluating
partition function. But the present treatment has shown the rate
equation in a general formulation of bimolecular reactions, and
thus has given it a much wider applicability. The formalism provides
a framework in terms of which molecular reactions can be understood
in a qualitative way. The kinetic theory values are too high
for all except atom$-$atom reactions. Hence, the transition-state
theory values can be regarded as exerting important control over
the rates of molecular reactions. It might be due to the high-pressure
limit that has led $k_{TS}$ to much closer agreement with experiment.

\newpage

\appendix

\noindent{\LARGE Addendum: Green's Function Method of Diffusion}

\vspace*{20pt}

Fick's diffusion equation suggests using Green's function
associated with the given partial differential equation.
Actually we can find in a mathematical physics book [4] a discussion
of the Green's function for diffusion. But the book has very
general discussion of the Green's function method. We must get
down to practical cases. In this Addendum, we shall work through
simple examples to see how it goes.

We consider hydrodynamic derivation on an intuitive basis of
the diffusion equation. Let $\rho({\bf r}, t)$ be the mean number
of molecules per unit volume located at time $t$ near the position
{\bf r}. If we assume that the total number of molecules $N$ in a
volume $V$ is conserved, we have, on account of the constancy
of number,
\begin{equation}
\frac{dN}{dt}=0, \quad\mbox{or}\quad
\frac{1}{\rho}\frac{d\rho}{dt}+\frac{1}{V}\frac{dV}{dt}=0.
\end{equation}
In hydrodynamics, the rate of dilatation of the fluid is equal to the
divergence of velocity [5]. If $\nabla\cdot{\bf v}=0$, the size
of the volume element remains constant with time, that is, the fluid
is incompressible. Equation (A.1) is thus written
\begin{equation}
\frac{1}{\rho}\frac{d\rho}{dt}+\nabla\cdot{\bf v}=0.
\end{equation}
Using the convective time derivative, we get the continuity equation
in the familiar form. As a result, we can see the correspondence:
\begin{equation}
\frac{dN}{dt}=0 \Longleftrightarrow
\frac{\partial\rho}{\partial t}+\nabla\cdot{\bf J}=0.
\end{equation}
When we use the diffusion equation ${\bf J}=-D\nabla\rho$,
we get the partial differential equation, satisfied by $\rho$. In the
simple case of a dilute fluid, it gives
\begin{equation}
\frac{\partial\rho}{\partial t}-D\nabla^2\rho=0.
\end{equation}

If there is a source or sink of particles, particles are not conserved
and (A.1) becomes inhomogeneous equation that has
a source or sink term,
\begin{equation}
\frac{dN}{dt}=-\lambda N.
\end{equation}
Hence
\begin{equation}
\frac{\partial\rho}{\partial t}-D\nabla^2\rho=-\lambda\rho.
\end{equation}
The form of the diffusion equation suggests using the Green's function
associated with the diffusion equation for its solution.
We need to construct the
corresponding time-dependent Green's function that satisfies
\begin{equation}
-\frac{\partial G}{\partial t}-D\nabla^2G=
-\delta({\bf r}-{\bf r}')\delta(t-t').
\end{equation}
The Green's function is
\begin{equation}
G({\bf r},t;{\bf r}',t')=
\frac{e^{-({\bf r}-{\bf r}')^2/4D(t-t')}}{[4\pi D(t-t')]^{3/2}}.
\end{equation}
To interpret (A.7) let $G$ be the density of molecules in a medium.
Then the impulsive point source involves introducing a unit of molecules
at ${\bf r}'$ at time $t'$. The Green's function then gives the density
at a future time for any other point of the medium and thus describes
the manner in which molecules diffuse away from its initial position.
It is a function of only the relative distance and the relative time
between source and observation point. As in the case of the scalar
wave equation, particular integral of the inhomogeneous diffusion
equation (A.6) is
\begin{equation}
\rho({\bf r},t)=\int G({\bf r},t;{\bf r}',t')\lambda\rho({\bf r}',t')
\,d^3r'\,dt'.
\end{equation}
By differentiating with respect to $({\bf r},t)$, we obtain
\begin{eqnarray}
\biggl(\frac{\partial}{\partial t}-D\nabla^2 \biggr)\rho({\bf r},t)
&=&\biggl(\frac{\partial}{\partial t}-D\nabla^2 \biggr)
\int G({\bf r},t;{\bf r}',t')\lambda\rho({\bf r}',t')
\,d^3r'\,dt'\\[2ex]
&=&\int\biggl(-\frac{\partial}{\partial t'}-D\nabla'^2 \biggr)
G({\bf r},t;{\bf r}',t')\lambda\rho({\bf r}',t')\,d^3r'\,dt',\nonumber
\end{eqnarray}
where uses have been made of $\nabla G=-\nabla' G$ and
$\partial G/\partial t=-\partial G/\partial t'$.
Formal interpretation of (A.7) may be obtained by considering the change
of differential variables.

The diffusion equation differs in many qualitative aspects from the
scalar wave equation, and of course the Green's function will exhibit
these differences. The most important single feature is the asymmetry
of the diffusion equation with respect to the time variable. Multiplying
(A.6) by $G$ and (A.7) by $\rho$, subtracting the two equations, and
integrating over space and over time gives
\begin{equation}
\int\biggl(G\frac{\partial\rho}{\partial t'}+
\rho\frac{\partial G}{\partial t'}\biggr)\,d^3r'\,dt'+
\int D\bigl(\rho\nabla'^2G-G\nabla'^2\rho\bigr)
\,d^3r'\,dt'=-\int G\lambda\rho\,d^3r'\,dt'+\rho({\bf r},t).
\end{equation}
We may apply the Green's theorem to the second of these integrals.
In the case of the first, the time integration may be performed.
Finally
\begin{equation}
\int G\rho \,d^3r'+
\int D\biggl(G\frac{\partial \rho}{\partial n}-
\rho\frac{\partial G}{\partial n}\biggr)\,dS\,dt'+
\int G\lambda\rho\,d^3r'\,dt'=\rho({\bf r},t),
\end{equation}
where $\partial/\partial n$ is the normal derivative at the surface
$S$ directed outward from inside the volume defined by $d^3r'$.
$G$ is chosen so as to satisfy homogeneous boundary conditions
corresponding to the boundary conditions satisfied by $\rho$.
While the first term of (A.12) includes the effects of the initial
value of $\rho$, other two terms represent the familiar effects
of volume sources and boundary conditions. The various inhomogeneous
equations and the initial-value problem can be solved in terms of
the Green's function which satisfies homogeneous boundary conditions
and a causality condition.

The problem discussed thus far is that of the absorption of one
substance by another through which it can diffuse. This can
be regarded as a problem in diffusion in which some of the diffusing
substance becomes immobilized as diffusion proceeds, or as a
problem in chemical kinetics in which the rate of reaction depends
on the rate of supply of one of the reactants by diffusion. Chemical
reactions are often considerable dependent on the mobility of the
reactants as well as on the kinetics of the reaction itself. They are
practical examples of processes involving simultaneous diffusion
and reaction of one sort or another [6].

An irreversible first-order reaction will be an illustration.
The rate of change of the total number of reactants is
\begin{equation}
\frac{dN}{dt}=-kN.
\end{equation}
By writing in terms of concentration, we obtain
\begin{equation}
\frac{\partial\rho}{\partial t}-D\nabla^2\rho=-k\rho,
\end{equation}
and we get the solution in the form of (A.12), with the rate
constant $k$ replacing $\lambda$. Near-exact form of (A.14)
has been used in the problems of conduction of heat and of
carrier concentration in semiconductor.

For an irreversible second-order reaction, the rate equation is
\begin{equation}
\frac{dN}{dt}=-kN^2.
\end{equation}
The procedure (A.1)$-$(A.4) gives for second-order kinetics
\begin{equation}
\frac{\partial\rho}{\partial t}-D\nabla^2\rho=-kN\rho.
\end{equation}
In form this is ``integrodifferential'' equation. But we can reduce
this to a differential equation. The total number $N$ is a function
of time. It is not a function of time and coordinates.
The time derivative of $N$
in (A.15) is differentiation following the motion of reactants. Thus
the solution of (A.15) may be used to give
\begin{equation}
\frac{\partial\rho}{\partial t}-D\nabla^2\rho=
-k\biggl(\frac{N_0}{1+N_0kt}\biggr)\rho.
\end{equation}
We may then expect the solution in the form of (A.12), but now with
$kN_0/(1+N_0kt)$ replacing $\lambda$. Basically what we have done
is to use the solution of the reaction rate equation to construct a
solution of inhomogeneous diffusion equation. It shows how the general
form of equations can be deduced by using the solutions of the
corresponding kinetics in the resulting diffusion equation.

Usually more than one reaction contributes to the rate of change of a
given species [7]. Organic reactions occur almost by a chain of mechanism.
But once the resulting rate equation is given, its formal solution is
obtained in integral form by use of the Green's theorem. The Green's
function method shows a mathematical power of solving the problem of
differential equation in terms of the integral equation formulation.
One can see at once that it includes explicitly terms that depend
upon the boundary conditions and transport properties of the reactant
molecules. Thus, it is likely to be much better than the reaction rate
equation, which provides no way to understand the transport process
in a system in which the chemical reaction occurs.


\begin{thebibliography}{0}

\bibitem{1} K. J. Laidler, {\it Chemical Kinetics} (Harper \& Row, 1987),
3rd ed.
\bibitem{2} F. Reif, {\it Fundamentals of Statistical and Thermal Physics}
(McGraw-Hill, 1965); R. H. Fowler, {\it Statistical Mechanics}
(Cambridge, 1936), 2nd ed.
\bibitem{3} K. A. Holbrook, {\it Chem. Soc. Rev.} {\bf 12}, 163 (1983);
P. H\"{a}nggi, P. Talkner, and M. Borkovec, {\it Rev. Mod. Phys.} {\bf 62},
251 (1990).
\bibitem{4} P. M. Morse and H. Feshbach, {\it Methods of Theoretical
Physics} (McGraw-Hill, 1953), Part 1.
\bibitem{5} A. G. Webster, {\it Dynamics of Particles and of Rigid,
Elastic, and Fluid Bodies} (Stechert-Hafner, 1920).
\bibitem{6} J. Crank, {\it Mathematics of Diffusion} (Oxford, 1975),
2nd ed.
\bibitem{7} W. J. Moore, {\it Physical Chemistry} (Prentice-Hall, 1972),
4th ed.; G. W. Castellan, {\it Physical Chemistry}
(Benjamin/Cummings, 1983), 3rd ed.

\end{thebibliography}
\end{document}